\documentclass[%
reprint,
twocolumn,
superscriptaddress
]{revtex4-1}
\usepackage{amsmath}
\usepackage{mathtools} 
\usepackage{graphicx}
\usepackage{dcolumn}
\usepackage{bm}
\usepackage{hyperref}
\hypersetup{
	bookmarks=true,
	colorlinks=true,
	urlcolor=blue,
	citecolor=blue,
	linkcolor=blue, 
	pdftex,
	linktocpage=true, 
	linktoc=all,     
	hyperindex=true
}
\usepackage[mathlines]{lineno}

\usepackage{siunitx} 
\usepackage{comment} 
\usepackage{braket} 

\begin{document}


\title{The Fate of the Higgs Mode in a Trapped Dipolar Supersolid}%

\author{J. Hertkorn}
\affiliation{5. Physikalisches Institut and Center for Integrated Quantum Science and Technology, Universität Stuttgart, Pfaffenwaldring 57, 70569 Stuttgart, Germany
}%
\author{F. B\"ottcher}%
\affiliation{5. Physikalisches Institut and Center for Integrated Quantum Science and Technology, Universität Stuttgart, Pfaffenwaldring 57, 70569 Stuttgart, Germany
}%
\author{M. Guo}%
\affiliation{5. Physikalisches Institut and Center for Integrated Quantum Science and Technology, Universität Stuttgart, Pfaffenwaldring 57, 70569 Stuttgart, Germany
}%
\author{J.N. Schmidt}%
\affiliation{5. Physikalisches Institut and Center for Integrated Quantum Science and Technology, Universität Stuttgart, Pfaffenwaldring 57, 70569 Stuttgart, Germany
}%
\author{T. Langen}%
\affiliation{5. Physikalisches Institut and Center for Integrated Quantum Science and Technology, Universität Stuttgart, Pfaffenwaldring 57, 70569 Stuttgart, Germany
}%
\author{H.P. B\"uchler}%
\affiliation{Institute for Theoretical Physics III and Center for Integrated Quantum Science and Technology, Universität Stuttgart, Pfaffenwaldring 57, 70569 Stuttgart, Germany}
\author{T. Pfau}%
\email{t.pfau@physik.uni-stuttgart.de}
\affiliation{5. Physikalisches Institut and Center for Integrated Quantum Science and Technology, Universität Stuttgart, Pfaffenwaldring 57, 70569 Stuttgart, Germany
}%

\date{\today}

\begin{abstract}
We theoretically investigate the spectrum of elementary excitations of a trapped dipolar quantum gas across the BEC-supersolid phase transition. Our calculations reveal the existence of distinct Higgs and Nambu-Goldstone modes that emerge from the softening roton modes of the dipolar BEC at the phase transition point. On the supersolid side of the transition, the energy of the Higgs mode increases rapidly, leading to a strong coupling to higher-lying modes. Our study highlights how the symmetry-breaking nature of the supersolid state translates to finite-size systems.
\end{abstract}

\maketitle

Superfluids that spontaneously break $U(1)$ symmetries occur in many fields of physics ranging from condensed matter to high energy physics to cosmology. Under certain conditions, such superfluids can undergo an additional quantum phase transition, breaking a continuous translational symmetry leading to the counter-intuitive state of matter known as supersolid \cite{Leggett1970,Prokofev2007,Boninsegni2012}. While the experimental proof of this state in helium has remained elusive \cite{Chan2013}, it has been induced as a density modulation via the invocation with light fields in spin-orbit coupled Bose-Einstein condensates (BECs) \cite{Li2017} and BECs coupled to crossed optical cavities \cite{Leonard2017supersolid,Leonard2017higgsgoldstone}. In contrast to these experiments, recent work on arrays of dipolar quantum droplets \cite{Roccuzzo2018,Tanzi2019,Bottcher2019,Chomaz2019,Tanzi2019compressional,Guo2019,Natale2019} has shown that the supersolid state of matter can arise purely from internal interactions of a quantum many-body system.  In general, the additional breaking of the continuous translational symmetry gives rise to two collective modes: a phonon like Nambu-Goldstone mode as well as an amplitude Higgs mode \cite{Nambu2015,Pekker2014}. In this letter, we study the behavior of the Higgs mode in a trapped system at the transition from the superfluid to a supersolid phase.

Starting with the pioneering work on the behavior of the amplitude Higgs mode in condensed matter systems \cite{Sooryakumar1981,Littlewood1982}, the study of this collective mode  has been a subject of intense theoretical \cite{Huber2007,Huber2008,Huber2009,Podolsky2011,Bruun2014} and experimental \cite{Bissbort2011,Endres2012,Sherman2015,Leonhard2016,Leonard2017higgsgoldstone,Behrle2018} investigation. In contrast to particle physics, where Lorentz invariance and particle-hole symmetry provide a stable  Higgs particle \cite{Higgs1964}, this collective mode is in general strongly damped due to its coupling to Goldstone modes. So far its observation required systems, where the effective low energy action exhibits an at least approximate particle-hole symmetry, yielding Lorentz-invariant equations of motion and a decoupling of the amplitude Higgs mode from the Nambu-Goldstone modes \cite{Huber2008,Pekker2014,Leonhard2016}. 

We start our analysis by describing the expected behavior of the collective low energy modes across the superfluid to supersolid quantum phase transition in the thermodynamic limit. In the superfluid regime, we expect a single Goldstone mode characterized by the superfluid sound velocity.  Then, the quantum phase transition is triggered by the softening of the roton, which eventually vanishes at the transition; at this critical point, the system exhibits three gapless modes. In the supersolid phase, two gapless modes remain and correspond to the coupled Goldstone mode between the superfluid phase mode associated with the $U(1)$ symmetry breaking and the phonon mode associated with the translational symmetry breaking. In turn, the amplitude Higgs mode of the translational symmetry breaking becomes gapped and strongly damped due to the absence of particle-hole symmetry. 

The focus of this manuscript is on the modification of this scenario in trapped systems, where modes and their energies are discretized due to the finite size. We map out the collective excitation spectrum across the superfluid-supersolid phase transition in a trapped system as experimentally realized with dipolar atomic gases \cite{Tanzi2019,Bottcher2019,Chomaz2019,Tanzi2019compressional,Guo2019,Natale2019} and identify how both the Goldstone and Higgs modes emerge. The analysis is based on the numerical investigation of the Gross-Pitaevskii equation (GPE) and the Bogoliubov-de Gennes (BdG) equations, which is a well established method to describe the quantum phase transition as well as the low-lying collective modes.  We identify a characteristic behavior of the Nambu-Goldstone mode and the amplitude Higgs mode close to the phase transition. Remarkably,  the amplitude Higgs mode is stable close to the quantum phase transition even in absence of particle-hole symmetry due to the energetic splitting of the modes in finite-size systems of trapped geometries.  Away from the critical point, we find that the increasing gap of the amplitude Higgs mode, then leads to a strong hybridization and avoided level crossings, which is the corresponding phenomenon to the strong damping of the amplitude Higgs mode in the thermodynamic limit. We emphasize that this characteristic fingerprint of the collective modes can be used as a tool to identify the precise position of the quantum phase transition in trapped systems and will provide new incentives to experimentally discover the Higgs mode in a system, where it arises purely from intrinsic interactions.

In view of the recent discovery of supersolidity in ultracold dipolar atomic gases \cite{Roccuzzo2018,Tanzi2019,Bottcher2019,Chomaz2019,Tanzi2019compressional,Guo2019,Natale2019}, we consider a system of $N = 30\, 000$ $^{162}\mathrm{Dy}$ atoms, in an harmonic potential with trapping frequencies $\omega_\mathrm{trap} = 2\pi\, (30,\, 90,\, 110)\, \mathrm{Hz}$, where the magnetic field is aligned along the $y$-direction. It is the same system we considered in our previous work \cite{Guo2019}, where we investigated in detail the low-energy Goldstone mode and proved the system's supersolidity, including phase rigidity. 

\begin{figure}[tb!]
	\includegraphics[trim=0 0 0 0,clip,scale=0.56]{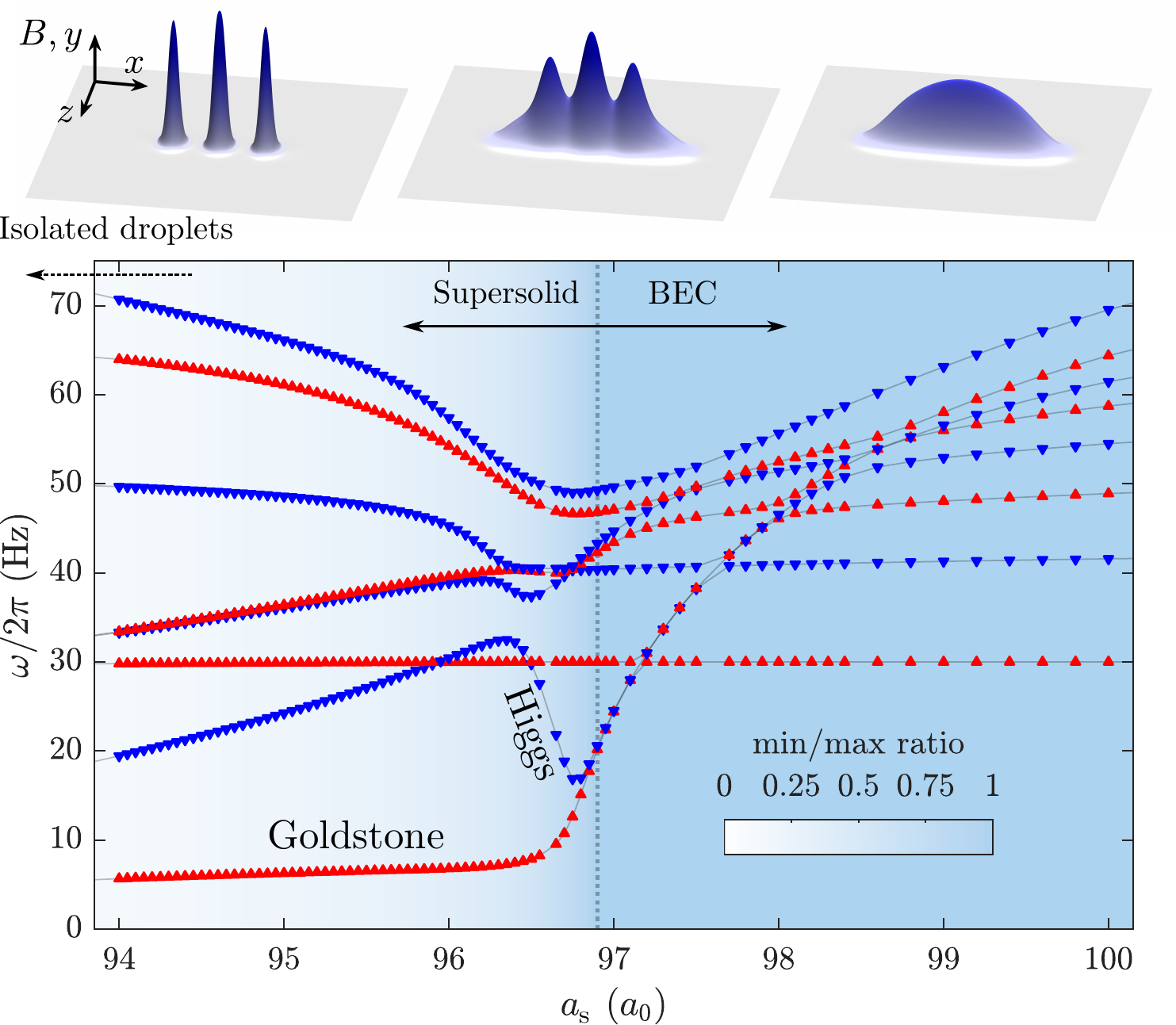}
	\caption{The excitation frequencies $\omega/2\pi$ of the eight lowest BdG modes as a function of the scattering length $a_\mathrm{s}$. Red (blue) triangles with a side facing up (down) indicate that the density variation function of the corresponding mode has odd (even) parity. The background color indicates the ratio between density in between central and side droplets and peak density of the center droplet, as a simple measure of the superfluid fraction \cite{Bottcher2019,Guo2019}. The vertical dashed line indicates the position of the phase transition. On top, we show density profiles in the $x$-$z$-plane for the isolated droplets (left), coherent supersolid droplets (middle) and BEC (right) regime.}
	\label{fig:spectrum}
\end{figure}

Such a system is described by the extended Gross-Pitaevskii equation (eGPE) \cite{Ronen2006,Wenzel2017,Roccuzzo2018,supmat}
\begin{equation}\label{eq:GPE}
	i \hbar \partial_t \psi = H_\mathrm{GP} \psi,
\end{equation}
where we define $H_\mathrm{GP} \coloneqq H_0 + g|\psi|^2 + \Phi_\mathrm{dip}[\psi] + g_\mathrm{qf} |\psi|^3$ and $\psi$ is normalized to the atom number $N=\int \mathrm{d}^3r\, |\psi(\boldsymbol{r})|^2$. $H_0 = \hbar^2 \nabla^2 / 2m + V_\mathrm{ext}$ contains the kinetic energy and trap confinement $V_\mathrm{ext}(\boldsymbol{r}) = m(\omega_x^2 x^2 + \omega_y^2 y^2 + \omega_z^2 z^2)/2$.  The contact interaction strength $g = 4\pi\hbar^2a_\mathrm{s}/m$ is given by the scattering length $a_\mathrm{s}$ and the dipolar mean field potential $\Phi_\mathrm{dip}$ by the dipolar length $a_\mathrm{dd} = \mu_0 \mu_\mathrm{m}^2 m / (12 \pi \hbar^2)$ ($a_\mathrm{dd}\approx 130\, a_0\ \text{for } ^{162}\mathrm{Dy})$, where $\mu_\mathrm{m}$ is the magnetic moment. Within the local density approximation, quantum fluctuations are taken into account by the term $g_\mathrm{qf}|\psi|^3$, which is the Lee-Huang-Yang (LHY) correction \cite{Lee1957} for dipolar systems \cite{Lima2011,Lima2012}, necessary to prevent the dipolar collapse \cite{Petrov2015,Kadau2016}. First, we perform imaginary time evolution of Eq.~(\ref{eq:GPE}) to find the ground state $\psi_0$. We obtain, as a function of the scattering length  a regular BEC, a supersolid or an array of isolated droplets, as schematically shown at the top of Fig.~\ref{fig:spectrum}.

In order to study the elementary excitations, we follow Refs.~\cite{Morgan1998,Huepe2003,Ronen2007,Baillie2017,Chomaz2018,Roccuzzo2018} and shortly review the BdG equations here. We linearize the wavefunction $\psi(\boldsymbol{r},t) = [\psi_0(\boldsymbol{r}) + \lambda (u(\boldsymbol{r}) e^{-i\omega t} + v^*(\boldsymbol{r})e^{i\omega t})]e^{-i\mu t / \hbar}$ around its ground state $\psi_0$ with the Bogoliubov amplitudes $u$ and $v$, oscillating at a frequency $\omega/2\pi$ as a small pertubation with amplitude $\lambda$.  $\mu$ is the chemical potential and $u,\ v$ are normalized by $\int \mathrm{d}^3r \, (u(\boldsymbol{r})^2 - v(\boldsymbol{r})^2) = 1$. Substituting the expression for $\psi$ in Eq.~\ref{eq:GPE} and retaining terms up to linear order in $\lambda$ yields a system of linear equations that can be neatly expressed in matrix form
\begin{equation}\label{eq:BdG}
	\begin{pmatrix}
		H_\mathrm{GP}-\mu+\hat{X} & \hat{X} \\
		-\hat{X} & -(H_\mathrm{GP}-\mu+\hat{X})
	\end{pmatrix}
	\begin{pmatrix}
	u \\
	v
	\end{pmatrix}
	= \hbar \omega 
	\begin{pmatrix}
	u \\
	v
	\end{pmatrix},
\end{equation}
where the operator $\hat{X}$ \cite{Baillie2017,Chomaz2018} is defined by its action on either function $\chi=u,\,v$ in the supplementary material \cite{supmat}. $\hat{X}$ contains the contact and dipolar interactions, as well as quantum fluctuations. It is essential to note here that our system intrinsically requires all three contributions, as a BEC with only contact and dipolar interactions in our parameter region of scattering lengths would collapse. Only through the stabilization with beyond mean-field effects given by the LHY correction, the droplet phase can exist \cite{Petrov2015,Kadau2016,Ferrier-Barbut2016}.

By solving Eq.~(\ref{eq:BdG}) for different scattering lengths across the BEC-supersolid phase transition, we gain insight into the energetic and spatial character of the elementary excitations in the self-organized supersolid regime. The excitation energies as a function of the scattering length are shown in Fig.~\ref{fig:spectrum}. Here we focus on the low-lying modes, in particular the eight lowest, which are decoupled from the $y$- and $z$-direction and mainly feature an excitation along the droplet array. It is instructive to label the energy branches by the corresponding mode's parity as even (odd), if the corresponding function that describes the density fluctuation is a symmetric (anti-symmetric) function along the direction of the droplet array \cite{supmat}.
The mode parity is essential in understanding the spectrum. As a function of one interaction parameter (in our case the scattering length $a_\mathrm{s}$), energy levels with modes of opposite parity are allowed to cross each other, while energy levels with modes of equal parity hybridize, leading to avoided crossings and level repulsion \cite{Neumann-Wigner1929, landau1981quantum}.

To distinguish the superfluid and supersolid phases, we look at the droplet density overlap, as defined in our previous works \cite{Bottcher2019,Guo2019} through the ratio of density in between central and side droplets and peak density in the central droplet, which is shown as the blue background in Fig.~\ref{fig:spectrum} (noted as min/max ratio). One can also think of this quantity as an inverse measure of the density modulation strength, which is connected to the superfluid fraction. In Fig.~\ref{fig:spectrum}, we indicate the phase transition by a dashed line for clarity. Let us now begin by considering the BEC phase at $a_\mathrm{s} = 100\,a_0$ on the right hand side of Fig.~\ref{fig:spectrum} and move in our discussion systematically towards smaller scattering lengths. 
The lowest branch is the anti-symmetric dipole mode, located at $30\, \mathrm{Hz}$, exactly the trap frequency. It is not affected by a change of scattering length, which is an instance of Kohn's theorem \cite{Khon1961} that demands the dipole mode to be always completely decoupled from all interactions, and thus from all other modes. This can serve as a test of the numerical accuracy of our calculations \cite{Ronen2006},
and our simulation recovers the dipole mode exactly for all considered scattering lengths. The next higher mode is the lowest quadrupole mode, which is followed by other modes, that have been studied for non-dipolar \cite{Stringari1996} and dipolar \cite{Odell2004, Bijnen2010} systems in various trap geometries. We find that the parity of the modes is alternating between odd and even as a function of the excitation energy on the BEC side, which energetically separates the modes well, as the kinetic energy increases in alternating parity. 


Moving towards smaller scattering lengths, one can recognize clearly that a bundle of branches quickly decreases in energy (see $a_\mathrm{s} \in (97.0, 98.5)\,a_0$ in Fig.~\ref{fig:spectrum}). It consists of the modes induced by the roton softening \cite{Giovanazzi2004,Corson2013,Chomaz2018} and is always composed of one symmetric and one anti-symmetric mode, which are degenerate until the scattering length approaches the phase transition (around $a_\mathrm{s} = 96.8\,a_0$). We note that the roton instability has previously been predominantly studied in 2D- systems \cite{Santos2003,Ronen2007,Wilson2008,Jona-Lasinio2013,Natu2014,Shen2018} and recently in 1D-systems \cite{Petter2018,Chomaz2018}. One can clearly see the anti-crossings between modes of equal parity as the roton mode softens. 


Further towards smaller scattering lengths, around $a_\mathrm{s} = 96.8\,a_0$, the system approaches the superfluid-supersolid phase transition, where these two modes continuously split in energy, losing their degeneracy. This can be intuitively understood, as, in contrast to the normal modes in the BEC that are defined by the confinement, the roton is defined by the intrinsic interactions and is localized only in the center of the BEC, where the density is almost flat. Even in our finite system, we remarkably observe that this leads to a degeneracy of the even and odd roton modes. Crossing the phase transition, a density modulation emerges in the ground state and one is no longer able to freely choose the position at which to place the modulation. This lifts the degeneracy of even and odd modes that match their wavelength to the spacing of the droplets. Thus, we emphasize that the significant change at the phase transition is the energetic splitting of the roton mode into the Goldstone and Higgs modes of the supersolid. In fact the splitting point of the roton bundle into Higgs and Goldstone branches may serve as a rigorous definition for the transition point from the superfluid to a supersolid state. It is interesting to note that this splitting point coincides exactly with the simple measure of overlap between the droplets that has been used in recent works \cite{Bottcher2019,Guo2019}, and is shown as the blue background in Fig.~\ref{fig:spectrum}. It underlines the validity of this simple measure to characterize the quantum critical point, as it was done in \cite{Bottcher2019,Guo2019}; see also \cite{Zhang2019}.

\begin{figure}[tb!]
	\includegraphics[trim=0 0 0 0,clip,scale=0.56]{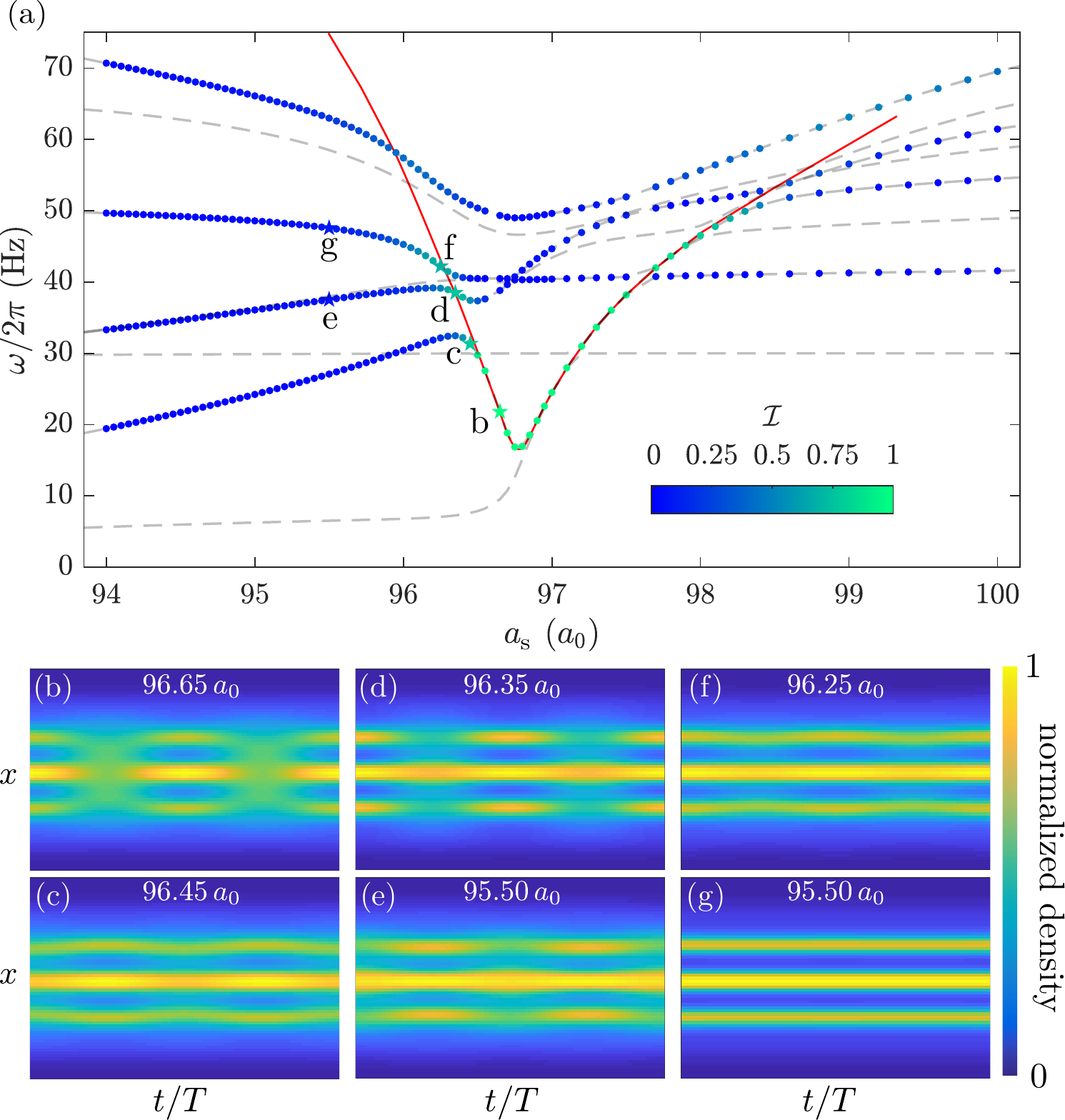}
	\caption{The upper panel (a) shows the points of Fig.~\ref{fig:spectrum} for symmetric modes, color coded with the Higgs coupling $\mathcal{I}$ to other modes. For clarity, the points of anti-symmetric modes have been removed, as the coupling to these modes is identically zero. The red line is a guide to the eye along the path of strongest Higgs character within the modes. The lower panel (b-g) shows time density evolution of the mode patterns at the selected points in (a) according to Eq.~\ref{eq:timeDensitySlice}, where (b) corresponds to $f^\mathrm{Higgs}$ and defines the coupling to all other modes. (b, d, f) show mode patterns at scattering lengths of optimal coupling on the (1st, 2nd, 3rd) symmetric branch. (c, e, g) show mode patterns at lower scattering lengths on the respective branches. One can see that the top row (b, d, f) shows the strong amplitude character, whereas the lower row (c, e, g) displays the mixing with breathing patterns and decay of the Higgs character within these respective branches. Every time axis has been rescaled to show 2 periods.}
	\label{fig:followinghiggs}
\end{figure}

This splitting of the roton bundle on the BEC side into the Higgs and the Goldstone mode of the supersolid can be seen as the main feature in Fig.~\ref{fig:spectrum}. The even mode that increases its energy with decreasing scattering length features an oscillation between the droplet crystal and the superfluid background and therefore can be identified as the Higgs mode (as illustrated further below, see Fig.~\ref{fig:followinghiggs}(b)). On the other hand, the decreasing mode is the low-energy Goldstone mode that has been recently described in \cite{Guo2019}. Despite the drastic change in energy, we observe that the mode pattern of these modes does not strongly change throughout the phase transition \cite{supmat}.

The Higgs mode in an infinite system is theoretically expected to be gapped, with a vanishing gap only exactly at the quantum critical point \cite{Huber2007,Huber2008,Podolsky2011,Pekker2014}. Due to the finite size of the here considered trapped system, the Higgs gap at the phase transition is lifted to a finite frequency of $\Delta_\mathrm{min} = 16.97\, \mathrm{Hz}$ at $a_\mathrm{s} = 96.75\, a_0$. In systems closer to the thermodynamic limit, such as with higher atom number and lower longitudinal confinement we expect the minimum Higgs gap at the quantum critical point to decrease, consistent with the predictions of a vanishing gap in the thermodynamic limit \cite{Huber2007,Huber2008,Huber2009,Podolsky2011,Pollet2012}.

Contrary to the gapped Higgs mode, the Goldstone mode in an infinite system is always gapless \cite{Huber2007,Huber2008,Huber2009,Podolsky2011,Bruun2014}. This again translates to a finite frequency of the Goldstone mode in our finite-size system, seen as the lowest anti-symmetric mode in Fig~\ref{fig:spectrum}. As the superfluid fraction decreases towards smaller scattering lengths, the frequency approaches zero, consistent with the decrease of the associated sound velocity of this mode. This is when the first symmetric mode also reaches energies below the dipole mode energy. We can identify this branch as the higher-$k$ Goldstone mode, with the lowest asymmetric branch being the $k=0$-Goldstone mode  (labeled "Goldstone" in Fig.~\ref{fig:spectrum}).

The phase transition from BEC to supersolid breaks the translational symmetry, in addition to the already broken $U(1)$ symmetry, which is essential for the existence of the Higgs (amplitude) mode \cite{Huber2008}. The broken translational symmetry means that the system is no longer Galilean invariant, thus we expect the Higgs mode to be overdamped \cite{Bissbort2011,Pollet2012,Podolsky2012,Pekker2014}. We illustrate this further in Fig.~\ref{fig:followinghiggs}, where we show the coupling strength of the Higgs mode to other modes and provide a view into the spatial modulation pattern. However, exactly at the phase transition, in a narrow range of scattering lengths $a_\mathrm{s} \in (96.45, 96.65)\,a_0$ the Higgs mode exists in an isolated state and is not coupled to any other mode, which is remarkable. This is a finite-size feature, since in the present system size, the energetic spacing between the lowest and other symmetric modes, especially at the quantum critical point, is significant.

To visualize how the Higgs mode couples to the other modes, we display the time evolution of these modes near the coupling points, as shown in Fig.~\ref{fig:followinghiggs}(b-g). The density time evolution \cite{Ronen2006,Martin2012} is given by
\begin{equation}\label{eq:timeDensitySlice}
\rho (x, t) = | \psi_0(x) |^2 + 2 \lambda f(x) \psi_0(x) \cos (\omega t),
\end{equation}
where $f(x) = u(x) + v(x)$ and $\psi_0(x)$ are 1D-cuts in the $y=0$ plane at $z=0$ through the density fluctuation $f$ and the ground state $\psi_0$, respectively.
Defining a measure on the strength of the Higgs character within each mode will allow us to follow its coupling and decay into the other symmetric modes more quantitatively. Here, we use the density mode pattern $f^\mathrm{Higgs}(x)$ at the scattering length $a_\mathrm{s} = 96.65\,a_0$ (Fig.~\ref{fig:followinghiggs}(b)) as a reference, since at this scattering length, the droplets have clearly developed, yet the energy of the Higgs branch is furthest from other symmetric modes, thus minimizing its coupling. We then calculate its fidelity (the modulus squared overlap)
\begin{equation}\label{eq:overlap}
\mathcal{I}[f] =\frac{\left|\braket{f|f^\mathrm{Higgs}}\right|^2}{\braket{f|f}\braket{f^\mathrm{Higgs}|f^\mathrm{Higgs}}},
\end{equation}
for all mode patterns $f$ and map it to a color code on the points shown in Fig.~\ref{fig:spectrum}, which results in the signature of Fig.~\ref{fig:followinghiggs}(a).

The time evolution of the Higgs density pattern (Fig.~\ref{fig:followinghiggs}(b)) shows that this mode is characterized by an oscillating exchange of amplitude between crystal and fluid, which is in line with the illustration of an amplitude oscillation with respect to the order parameter in the valley of the famous Mexican-hat potential \cite{Podolsky2011,Endres2012,Pekker2014,Guo2019,Leonard2017higgsgoldstone,Leonard2017supersolid}. For slightly lower scattering lengths, we observe that the branch on which we identify the most typical Higgs character now decreases in energy and develops an additional breathing character while maintaining part of the BEC background oscillation character, see Fig.~\ref{fig:followinghiggs}(c). As this occurs, the two higher lying symmetric modes increase in energy and develop the characteristic amplitude fluctuation of the Higgs mode, see Fig.~\ref{fig:followinghiggs}(d, f). In combination with an increase of the Higgs overlap (\ref{eq:overlap}) on these branches, we can identify that the actual Higgs branch steadily increases in energy away from the quantum critical point after its minimum has been surpassed, schematically indicated by the red line in Fig.~\ref{fig:followinghiggs}(a). Effectively, this means that the Higgs branch hybridizes with the other symmetric modes, as has been noted in a previous work considering strongly interacting lattice Bosons \cite{Bissbort2011}, and as a consequence decays into the other symmetric modes due to its overdamped nature \cite{Bissbort2011,Pollet2012,Podolsky2012,Pekker2014}. Fig.~\ref{fig:followinghiggs} shows remarkably well the coupling and decay of the Higgs mode to the higher symmetric modes, indicated by the green color, but also the loss of Higgs character within each branch. Following along the lowest symmetric branch in Fig.~\ref{fig:followinghiggs} towards smaller scattering lengths, the strength of the Higgs character within this branch decays rapidly, as it develops a more dominant breathing character, see Fig.~\ref{fig:followinghiggs}(c). With a further decrease of scattering length, the mode develops into the higher-$k$ Goldstone mode, that manifests as an oscillation of atoms between the central and the two side droplets \cite{supmat}. We additionally show points of strongest Higgs character within the two higher symmetric branches in Fig.~\ref{fig:followinghiggs}(d, f) and emergence of breathing character and decay along the branches in Fig.~\ref{fig:followinghiggs}(e, g). Towards the isolated droplet regime at even smaller scattering lengths, the Higgs mode has disappeared, as its character decays with the exponentially decreasing overlap between the droplets. We can relate this to the results obtained in two-dimensional superfluids \cite{Pollet2012,Endres2012,Bruun2014}, where the Higgs mode can only be seen close to the quantum critical point and its character disappears towards the Mott-insulating state of matter. 

In view of the existence of an isolated Higgs branch at the quantum critical point and its traceability into other symmetric modes within a range in scattering length of about $1\,a_0$, we propose the Higgs mode to be experimentally observable with current experimental setups \cite{Chomaz2018,Guo2019,Bottcher2019}. To this end, we have performed real time evolution simulations of the eGPE (\ref{eq:GPE}) using the standard experimental protocol of starting in the BEC regime with a subsequent linear ramp of the scattering length down into the supersolid state of matter. We systematically verified \cite{supmat} that in these simulations, the strongest density response is found to resemble the Higgs density pattern from our BdG calculations as together with a weaker density response that resembles the breathing patterns (as in the mixed state Fig.~\ref{fig:followinghiggs}(c)), proving the possibility of the experimental observation of the Higgs mode in the future.

In conclusion, we have mapped out the spectrum of collective excitations across the BEC-supersolid phase transition and identified the Goldstone and Higgs branches that emerge from the two roton modes. We thereby improve the understanding of the Goldstone mode, as measured in \cite{Guo2019}, and predict the existence of an isolated and stable Higgs mode in a narrow range of interaction strengths. The Higgs mode is gapped, its energy increases further away from the transition and it strongly couples to higher lying modes. Let us emphasize again: here we considered a system, exclusively subject to its own intrinsic interactions and a trapping potential. By varying only one of the intrinsic interaction parameters, the system undergoes a supefluid-supersolid phase transition. In the supersolid, the confinement enhances energetic splitting between Higgs and other modes, which leads to an isolated and decoupled Higgs mode. Due to the clear framework of the here considered system, it provides an excellent opportunity to study a correspondence between models in high energy physics \cite{Arkani-Hamed1998,Randall1999} and condensed matter systems.

As a next step, we plan to experimentally confirm the here presented findings and map out the coupling of the Higgs mode to higher lying modes.

\begin{acknowledgements}
	We acknowledge useful conversations with the participants F. Ferlaino, G. Modugno, L. Santos and T. Pohl's groups at the meeting on “Perspectives for supersolidity in dipolar droplet arrays” in Stuttgart as well as with R.M.V. Bijnen and J. Schmalian. We thank M. Wenzel for his contributions in the early stages of the project. This work is supported by the German Research Foundation (DFG) within FOR2247 under Pf381/16-1 and Bu2247/1, Pf381/20-1, and FUGG INST41/1056-1. T.L. acknowledges support from the EU within Horizon2020 Marie Sk\l odowska Curie IF (Grants No.~746525 coolDips).
\end{acknowledgements}




\bibliographystyle{apsrev4-1}
\bibliography{refs_cleaned} 

\clearpage


\section{Supplementary material}

\setcounter{figure}{0}

\renewcommand{\figurename}{Supplementary Figure}
\renewcommand{\thefigure}{S\arabic{figure}}

\subsection{Simulation details}
In this section, we quickly summarize simulation details not further explained in the main text. 

Our system is described by the eGPE Eq.~\eqref{eq:GPE}.
The dipolar mean field potential is given by
\begin{equation}\label{eq:meanfieldDipolar}
\Phi_\mathrm{dip}(\boldsymbol{r}) = \int \mathrm{d}^3r'\,  	V_\mathrm{dd}(\boldsymbol{r}-\boldsymbol{r}') |\psi(\boldsymbol{r}')|^2, 
\end{equation}
where 
\begin{equation}\label{eq:dipolarPotential}
V_\mathrm{dd}(\boldsymbol{r}) = \frac{3g_\mathrm{dd}}{4\pi} \frac{1-3\cos^2 \vartheta}{|\boldsymbol{r}|^3},
\end{equation}
is the dipolar interaction with the dipolar interaction parameter $g_\mathrm{dd} = 4\pi\hbar^2a_dd/m$ given by the dipolar length $a_\mathrm{dd} = \mu_0 \mu_\mathrm{m}^2 m / (12 \pi \hbar^2)$ defined by the magnetic moment $\mu_\mathrm{m}$ and $\vartheta$ is the angle between $\boldsymbol{r}$ and the magnetic field axis. 

Furthermore, $g_\mathrm{qf} = 32/(3\sqrt{\pi}) g a^{3/2} Q_5 (\epsilon_\mathrm{dd})$ is taking into account quantum fluctuations within the local density approximation for dipolar systems \cite{Lima2011,Lima2012}, where $\epsilon_\mathrm{dd} = g_\mathrm{dd} / g = a_\mathrm{dd} / a_\mathrm{s}$ is the relative dipolar strength. In our simulations, we use the approximation  $Q_5(\epsilon_\mathrm{dd}) = 1+ \frac{3}{2}\epsilon_\mathrm{dd}^2$ \cite{Lima2012,Ferrier-Barbut2016,Wenzel2017,Bisset2016,Baillie2017}. The mean field dipolar potential Eq.~(\ref{eq:meanfieldDipolar}) is effectively calculated using a Fourier transform, where we employ a spherical cutoff for the dipolar potential Eq.~(\ref{eq:dipolarPotential}), and choose its radius such that there is no spurious interaction between periodic images \cite{Goral2002,Ronen2006,Lu2010}. To find the ground state in all simulations, we use the method of imaginary time propagation.


In order to study the elementary excitations, we employ the BdG formalism introduced in the main text. The coupling operator $\hat{X}$ in the BdG matrix representation Eq.~(\ref{eq:BdG}) of the main text is given by \cite{Baillie2017,Chomaz2018}

\begin{widetext}
	\begin{equation}
	\hat{X}\chi(\boldsymbol{r}) = \psi_0(\boldsymbol{r}) \int \mathrm{d}^3r' [V_\mathrm{dd}(\boldsymbol{r}-\boldsymbol{r}') + g \delta(\boldsymbol{r}-\boldsymbol{r}')]\psi_0^*(\boldsymbol{r}')\chi(\boldsymbol{r}')
	+ \frac{3}{2} g_\mathrm{qf} |\psi_0(\boldsymbol{r})|^3 \chi(\boldsymbol{r})
	\end{equation}
\end{widetext}

Note that we follow the sign convention ($+u$, $+v$) \cite{Morgan1998,Ronen2006} in the expansion of $\psi$, instead of ($+u$, $-v$) \cite{Baillie2017,Chomaz2018}, which inverts the sign on the off-diagonal elements of the BdG matrix Eq.~(\ref{eq:BdG}). The particle (hole) components are usually, in our convention, identified by the functions $u$ ($v$) \cite{Ronen2006}. If we set $v\rightarrow-v$, we can convert the matrix representation used in our convention ($+u$, $+v$) to the matrix 
$$
\begin{pmatrix}
H_\mathrm{GP}-\mu+\hat{X} & -\hat{X} \\
\hat{X} & -(H_\mathrm{GP}-\mu+\hat{X})
\end{pmatrix}
$$
previously used in the ($+u$, $-v$) convention \cite{Baillie2017,Chomaz2018}.

In practice, we employ the same method described in \cite{Ronen2006}, which is to rewrite $f=u+v$ and $g = v-u$, corresponding to the density and phase fluctuations, respectively, which enables the reduction of Eq.~(\ref{eq:BdG}) after a short calculation into the equations 
\begin{equation}\label{eq:feq}
	(H_\mathrm{GP} - \mu)(H_\mathrm{GP} - \mu + 2X)f = \hbar^2 \omega^2 f
\end{equation}
and 
\begin{equation}\label{eq:geq}
(H_\mathrm{GP} - \mu + 2X)(H_\mathrm{GP} - \mu)g = \hbar^2  \omega^2 g.
\end{equation}
We solve the equation for $f$ and calculate the respective $g$ as in \cite{Ronen2006}. With the solutions for $f$ and $g$, $u = (f-g)/2$ and $v = (f+g)/2$ can be recovered. The solution for the equation of $f$ is obtained by using the Arnoldi method to calculate the lowest eigenvalues and eigenfunctions \cite{Huepe2003,Ronen2006,Martin2012}. Note that the BdG equations \eqref{eq:BdG} as presented in the main text always have the solution $g=\psi_0$ with $\omega = 0$, as one can easily verify from Eq.~\eqref{eq:geq}. This solution is sometimes referred to as neutral mode \cite{Morgan1998,Huepe2003,Ronen2006} and has its origin in fixing the phase of the wavefunction by introducing $\mu$ in the linearization. However, we emphasize that this is not a physical degree of freedom and refrain from calling it a "mode" since $(H_\mathrm{GP} - \mu)g = 0$ for $g = \psi_0$ is merely stating that the ground state fulfills the eGPE, which is redundant.

We point out that the conventional BdG description, as we present it in this paper, is linearizing the the wavefunction around its ground state with quasiparticles that are not orthogonal to the condensate \cite{Morgan1998}. In a previous study on nonlinear mixing of these quasiparticles \cite{Morgan1998}, it was shown that transfer of the quasiparticle population from one mode into another is inversely dependent on the the inital finite amount of excitation. The mixing processes are enhanced by a finite temperature and dominated by energy conserving conversion processes \cite{Morgan1998}. This will provide an additional theoretical challenge in the study of the coupling of the Higgs mode and its dependence with considerations of finite-temperature effects. In particular, a detailed study of the Higgs mode in the coupled regime and at higher energies may require an orthogonal quasiparticle basis.

We classify the excitation modes in terms of the parity of the density response. For this, we choose $\epsilon = 0.1$ as the criterion that determines the parity,
\begin{equation}\label{eq:parity}
\mathcal{P} = 
\begin{cases}
1, &\underset{x}{\max}\ q(x) > 1 - \epsilon\\
0, &\underset{x}{\max}\ q(x) < \epsilon
\end{cases},
\end{equation}
where $q(x) = |f(x)-f(-x)|/(2\, \underset{x}{\max}\ |f(x)|)$, $\mathcal{P} = 1,\, 0$ for odd or even modes, respectively.

\subsection{Elementary excitation spectra}
\begin{figure}[b!]
	\includegraphics[trim=0 0 0 0,clip,scale=0.3]{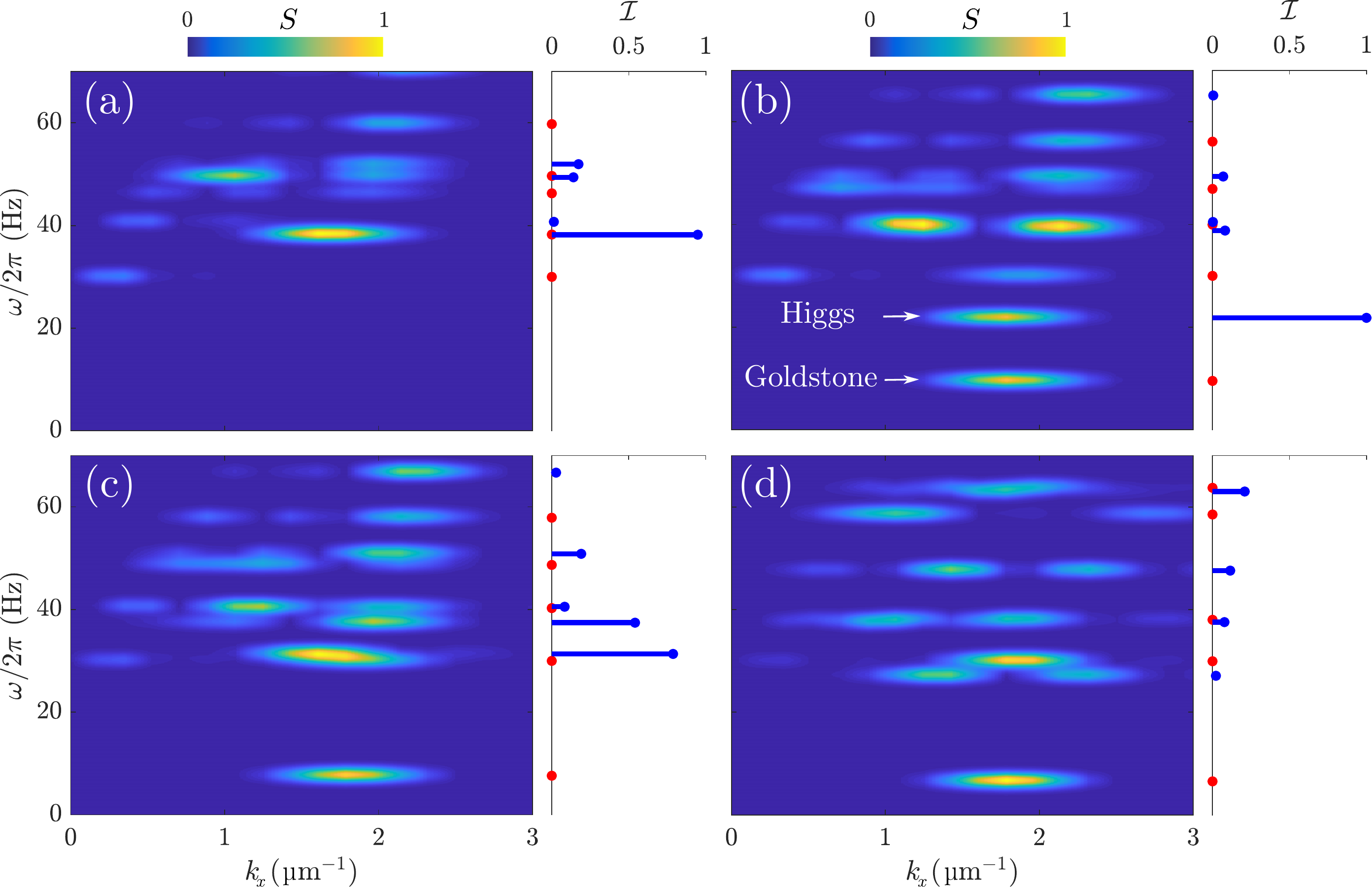}
	\caption{Structure factor spectrum (\ref{eq:SF}) for different scattering lengths $a_\mathrm{s} = \{ 97.50,\,96.65,\,96.45,\,95.50\}\, a_0$ in (a,$\,$b,$\,$c,$\,$d), respectively as color plots. The color code is normalized to the maximum structure factor across (a-d). Blue (red) bars on the right hand side of (a-d) indicate the Higgs strength $\mathcal{I}$ of every even (odd) mode at the corresponding energies.}
	\label{fig:SF}
\end{figure}

Since we have a finite sized system, we obtain a discrete spectrum of elementary excitations for every scattering length, rather than a continuous one for an infinite system. In order to visualize the momentum dependence of the obtained modes (which we will label by an index $j$ in the following), we calculate the zero-temperature dynamic strucure factor \cite{Brunello2001,Blakie2002}, given by 
\begin{equation}\label{eq:SF}
S(\boldsymbol{k}, \omega) = \sum_j \left| \int \mathrm{d}^3r\, (u_j^*(\boldsymbol{r})+v_j^*(\boldsymbol{r}))e^{i \boldsymbol{k}\cdot \boldsymbol{r}} \psi_0(\boldsymbol{r})\right|^2 \delta(\omega - \omega_j).
\end{equation}
It is also known as the density correlator \cite{Huber2007}, typically measured as a response function for Bragg spectroscopy, since it correlates and Fourier transforms the density fluctuation $f=u+v$ with the ground state $\psi_0$ as $\mathcal{F}[f^*\cdot \psi_0](\boldsymbol{k})$, where $\mathcal{F}$ is the Fourier transform. For illustration purposes in Fig.~\ref{fig:SF}, we convolve (\ref{eq:SF}) with a Gaussian along the $\omega$-axis.

In Fig.~\ref{fig:SF}, we show 4 different scattering lengths, $a_\mathrm{s} = \{ 97.50,\,96.65,\,96.45,\,95.50\}\, a_0$ corresponding to a BEC state close to the supersolid phase transition, a state that displays the isolated Higgs mode, a state that displays the mixed Higgs mode, and a state further towards the isolated droplet regime, respectively. The left hand panels show the dynamic structure factor Eq.~(\ref{eq:SF}), and the right hand side shows Higgs character $\mathcal{I}$ (Eq.~\ref{eq:overlap} in the main text), of the modes at the correponding energies. In Fig.~\ref{fig:SF}(a), we see a discrete version of the typical dispersion relation of dipolar BECs at an interaction strength that features the roton minimum. As the roton further softens towards smaller scattering lengths, the system approaches the phase transition and finally, as the quantum critical point has been crossed, splits into the Goldstone and Higgs modes, see Fig.~\ref{fig:SF}(b). The lowest mode is the low-energy Goldstone mode that we briefly discussed in the main text and was the main subject of our previous work \cite{Guo2019}. The symmetric mode above the Goldstone mode is the isolated Higgs mode, energetically well separated from other symmetric modes which all display a very small Higgs character $\mathcal{I}$, as one can see in the side panel of the spectrum. In contrast, at $a_\mathrm{s} = 96.45\, a_0$ (Fig.~\ref{fig:SF}(c)), as the Higgs mode rises in energy, it hybridizes with the higher-lying symmetric modes and mixes its character with them, resulting in a decreased Higgs strength on the lowest symmetric mode and an increased Higgs strength on the higher symmetric modes. Even further towards the isolated droplet regime (Fig.~\ref{fig:SF}(d)), the Higgs mode has disappeared (there is only residual overlap $ < 30\,\%$ with other symmetric modes).

We emphasize that the $k_x$ in Fig.~\ref{fig:SF}(b-c) is to be understood by folding the roton momentum $k_\mathrm{rot}$ back to zero, since after the density modulation has emerged, the dispersion relation of the system is best described within first Brillouin zone $k_x \in [0, k_\mathrm{rot})$. One can see in Fig.~\ref{fig:SF}(b) that the splitting of the roton mode into Goldstone and Higgs modes takes place at $k_\mathrm{rot}$ as well, so that the prominent isolated Higgs mode we mainly discuss in the main text is the $k=0$-Higgs mode. We observed the spectra such as the ones shown in Fig.~\ref{fig:SF} for all scattering lengths and checked for the $k$-dependence of the Higgs mode. In order to do so, we weighted every mode's structure factor with its Higgs overlap, and superimposed the weighted structure factors of all scattering lengths,
\begin{equation}
	\tilde{S}(\boldsymbol{k}, \omega) = \sum_{a_\mathrm{s} = 94.00\, a_0}^{a_\mathrm{s} = 100\, a_0} \mathcal{I}[f_{a_\mathrm{s}}](\omega) S_{a_\mathrm{s}}(\boldsymbol{k}, \omega),
\end{equation}
in order to filter out modes with smaller and keep modes with larger Higgs character. We find that the maximum strength of $\tilde{S}$ shifts slightly away from $k_\mathrm{rot}$, but exactly for the scattering lengths at which the Higgs mode hybridizes with the breathing modes. In this sense one cannot distinguish anymore whether $\tilde{S}$ probes partly the varying $k$ of the breathing modes or not.

\subsection{Experimental observability of the Higgs mode}
In order to verify that the isolated Higgs mode should be experimentally observable, we performed real time evolution simulations of the eGPE and followed a standard experimental protocol. Starting from a BEC ground state at $a_\mathrm{s} = 112\, a_0$, we linearly ramp the scattering length within $30\, \mathrm{ms}$ down into the supersolid regime at scattering lengths $a_\mathrm{s} = 96.65,\ 96.45\, a_0$ and observe the subsequent time evolution up to $300\, \mathrm{ms}$. 

\begin{figure}[tb!]
	\includegraphics[trim=0 0 0 0,clip,scale=0.4]{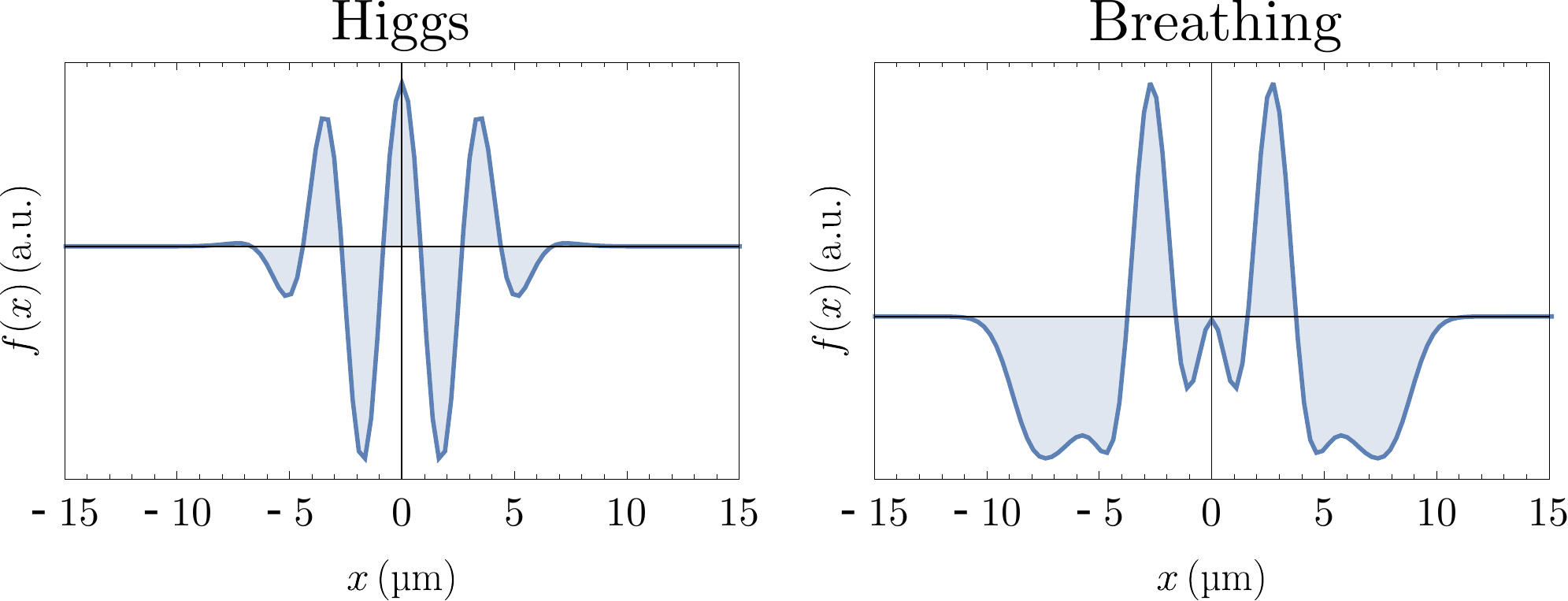}
	\caption{The theoretical mode patterns $f(x)$ from the BdG calculations at $a_\mathrm{s} = 96.65 a_0$ for the (1st, 3rd) lowest symmetric modes, corresponding to the (Higgs, breathing) patterns. The Higgs mode is characterized by a strong amplitude on the droplet positions (around $0,\, \pm 3.5\, \si{\micro}\mathrm{m}$), and opposite signed amplitude on the BEC background.}
	\label{fig:bdgModes}
\end{figure}

For later comparison, we show the BdG mode patterns corresponding to the pure Higgs and breathing density variation functions $f$ in Fig.~\ref{fig:bdgModes}. These are the 1st and 3rd lowest symmetric modes of Fig.~\ref{fig:spectrum} in the main text, at $a_\mathrm{s} = 96.65\, a_0$. The Higgs pattern corresponds to the time evolution shown in Fig.~\ref{fig:followinghiggs}(a). The breathing pattern is similar to Fig.~\ref{fig:followinghiggs}(c), but without a mixed Higgs character.

\begin{figure}[tb!]
	\includegraphics[trim=0 0 0 0,clip,scale=0.73]{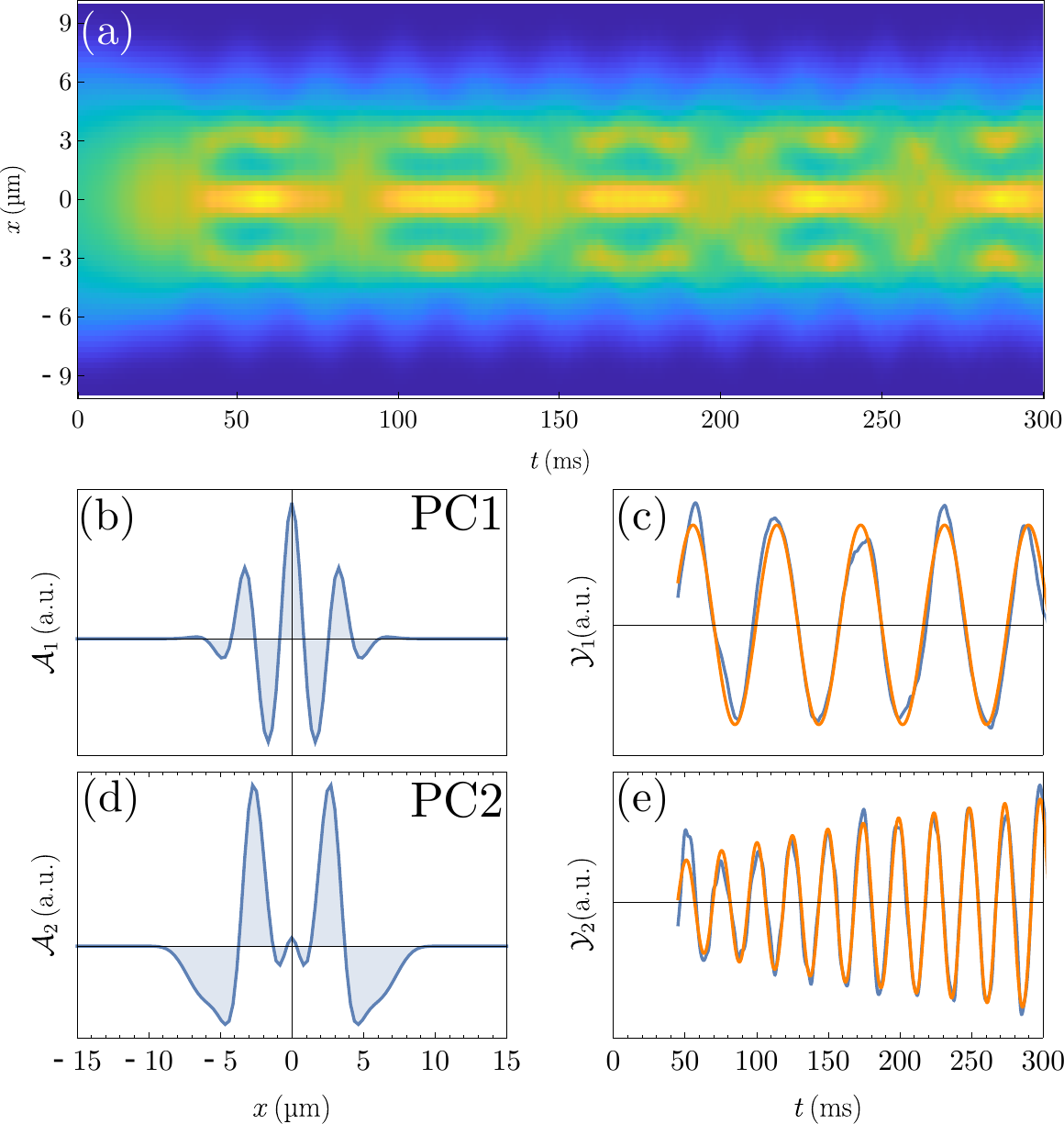}
	\caption{(a) shows the real time evolution of the experimental protocol to the final scattering length $a_\mathrm{s} = 96.65\, a_0$. The droplets have formed after approximately $t^* = 45\, \mathrm{ms}$. We perform PCA on the timeseries (a) for $t>t^*$ and show the (first, second) PC 1D-cut in panels (b,d). (c,e) show the weigths of the (first, second) PC as a function of time $t$ in blue and a fit in orange. (PC1, PC2) account for ($70.5\,\%$, $22.4\,\%$) of the total density variation contained in the timeseries (a). The frequencies of the PC's are extracted by fitting sinusoidal models to the weights and yield ($17.09$, $40.46$)$\pm 0.01\, \mathrm{Hz}$ for (PC1, PC2).}
	\label{fig:realTimeIsolatedHiggs}
\end{figure}

\begin{figure}[tb!]
	\includegraphics[trim=0 0 0 0,clip,scale=0.73]{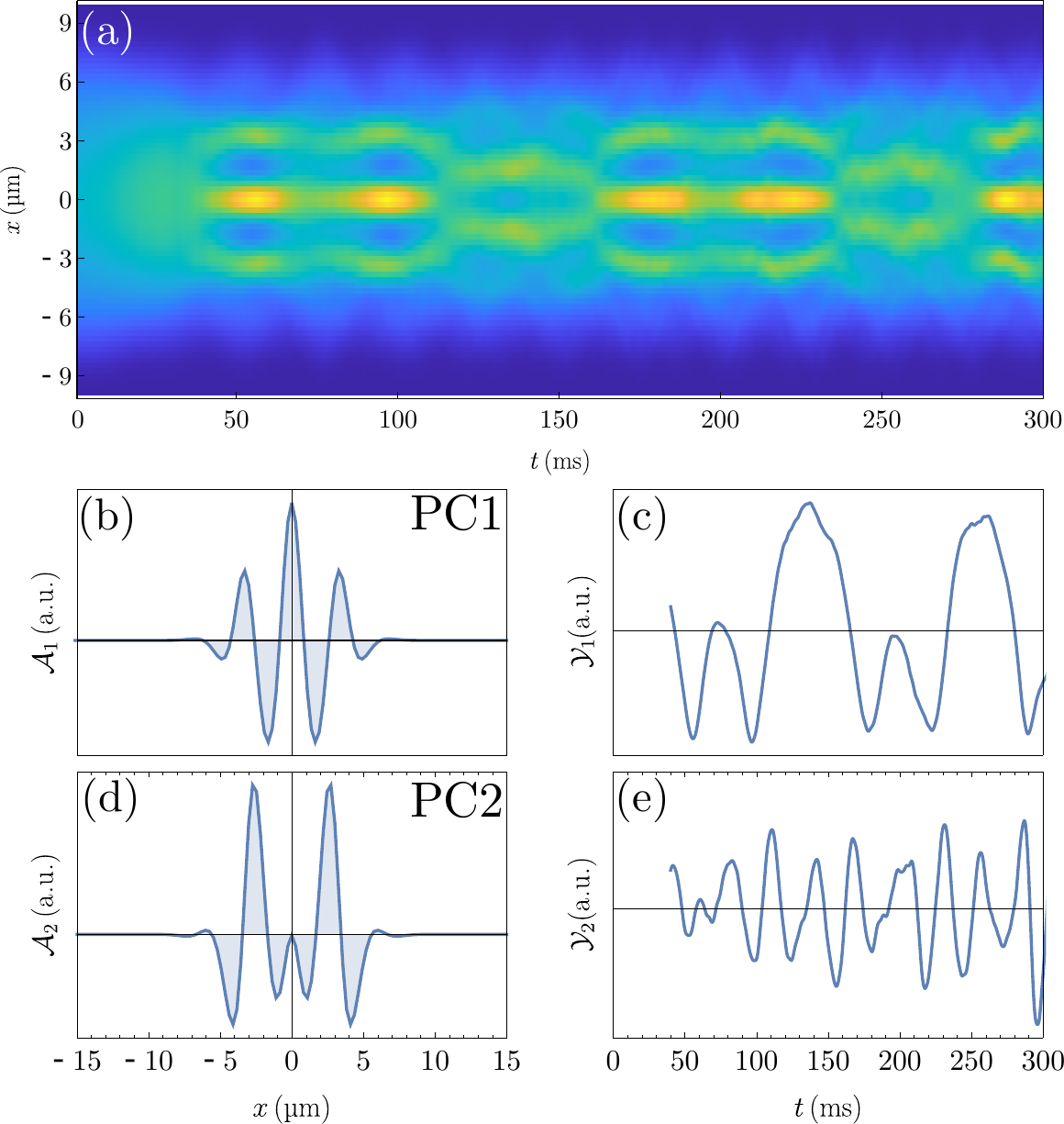}
	\caption{(a) shows the real time evolution of the experimental protocol to the final scattering length $a_\mathrm{s} = 96.45\, a_0$. The droplets have formed after approximately $t^* = 45\, \mathrm{ms}$. We perform PCA on the timeseries (a) for $t>t^*$ and show the (first, second) PC 1D-cut in panels (b,d). (c,e) show the weigths of the (first, second) PC as a function of time $t$. (PC1, PC2) account for ($86.1\,\%$, $10.1\,\%$) of the total density variation contained in the timeseries (a). We clearly see in the more complex time evolution of the weights, as compared to Fig.~\ref{fig:realTimeIsolatedHiggs}(c,e), that the obtained PC's do not account for a single mode, but rather that in the timeseries (a), Higgs and breathing modes are already coupled to further modes.}
	\label{fig:realTimeCoupledHiggs}
\end{figure}

By performing the experimental scheme for different final scattering lengths, we obtain the full density evolution shown in Figs.~\ref{fig:realTimeIsolatedHiggs},\,\ref{fig:realTimeCoupledHiggs} for the scattering length at which we expect to find an isolated Higgs mode, or a mode that displays mixed Higgs and breathing character, respectively. After the $30\, \mathrm{ms}$ ramp and another approximate $15\, \mathrm{ms}$, the droplets form and begin an oscillatory motion. 

In order to decompose this oscillatory motion into the elementary excitations, as we have studied them in the BdG formalism, we employ a model-free statistical analysis, the principal component analysis (PCA) \cite{Jolliffe2002,Segal2010,Bertram2014,Dubessy2014,Ferrier-Barbut2018,Cao2018}. In short, PCA provides a way to map a timeseries of density distributions to a set of density variation distributions, whose weight varies in time. PCA has a wide range of general applications \cite{Jolliffe2002}, from image analysis, to dimensional reduction of large datasets, but it turns out as well that there is a direct correspondence \cite{Dubessy2014,Barr2015} between the variation distributions, or principal components (PCs), and the density variation function $f$ obtained by the BdG formalism, so that we can identify a specific PC with a specific BdG mode and we may see the weight of this mode as a function of time. 
One of PCA's properties is that the signal (in our case the timeseries shown in Fig.~\ref{fig:realTimeIsolatedHiggs}(a),\,\ref{fig:realTimeCoupledHiggs}(a)) can be reconstructed exactly using a superposition of all PCs, oscillating at their respective weights. However, a very small subset out of all PCs accounts for most of the density variation (or information) contained within the timeseries. This is essentially how PCA is used for dimensional reduction. More explicitly stated, it means the following \cite{Segal2010,Dubessy2014}. Consider a set of $N_t$  images of $p$ pixels each. Typically for high resolution images or simulation grids, $N_t < p$, which we will assume in the following. We represent the $j$'th image by a vector $\boldsymbol{X}_j$, the mean image by $\boldsymbol{M} = \sum_{j=1}^{N_t} \boldsymbol{X}_j / N_t$, write the dataset in a $p \times N_t$ matrix $B = [\boldsymbol{X}_1 - \boldsymbol{M},\ ... \ , \boldsymbol{X}_{N_t} - \boldsymbol{M}]$ and apply PCA to this dataset. PCA diagonalizes the covariance matrix of $B$ and thus essentially finds a lower dimensional linear transform $Y$ (an $N_t \times N_t$ matrix) that represents the data $\{\boldsymbol{X}_j\}_{j=1}^{N_t}$ by a new set of $N_t$ orthonormal basis vectors $\{\boldsymbol{A}_j\}_{j=1}^{N_t}$ (the PCs), which are pairwise uncorrelated, sorted in decreasing variance and yield the original data exactly as
\begin{equation}\label{eq:PCAExactSum}
	\boldsymbol{X}_{j} = \boldsymbol{M} + \sum_{k=1}^{N_t} Y_{kj} \boldsymbol{A}_k, \quad j = 1,\ ..., \ N_t.
\end{equation}
PCA is a total-variance-preserving operation, meaning that the total variance of the signal is captured exactly by the sum of the variances of the individual PCs. But since we ordered the PCs in decreasing variance, it means that we can account for most of the variance, considering only a small subset $\{\boldsymbol{A}_j\}_{j=1}^{\tilde{N}}$ with $\tilde{N} \ll N$, of all PCs. We normalize the variance of each PC by the total variance as a relative measure of total variance accounted for. Reconstruction of the original dataset with only partial information is then written as
\begin{equation}\label{eq:PCAApproxSum}
\boldsymbol{X}_{j} \approx \boldsymbol{M} + \sum_{k=1}^{\tilde{N}} Y_{kj} \boldsymbol{A}_k, \quad j = 1,\ ..., \ N_t.
\end{equation}  

In the simulation, we take around $N_t = 1500$ snapshots of the wavefunction and we find that merely considering  $\tilde{N} = 2$ PCs, as we show further below, accounts for more than $90\,\%$ of the total variation.

From Eqs.~(\ref{eq:PCAExactSum},\,\ref{eq:PCAApproxSum}), we see that the weight of each PC for the $j$'th image is contained in the $j$'th column of $Y$. Hence, in the following we will label the 1D-cuts along the $x$-axis of the $k$'th PC as $\mathcal{A}_k(x)$, corresponding to $\boldsymbol{A}_k$ and its weight as a function of time as $\mathcal{Y}_k(t)$, corresponding to $\{Y_{kj}\}_{j=1}^{N_t}$.

In the comparison of the PCs shown in Figs.~\ref{fig:realTimeIsolatedHiggs},\,\ref{fig:realTimeCoupledHiggs}, the similarity between the BdG Higgs and breathing patterns shown in Fig.~\ref{fig:bdgModes} is very clear, so that we can identify that the Higgs mode has been clearly, and since it is the first PC, most strongly, excited for both values of the final scattering length $a_\mathrm{s} = 96.65,\ 96.45\, a_0$. In both cases we find the breathing pattern to be second most strongly excited and again we identify the strong similarity between the BdG mode pattern and the PCs. 

However, if we observe the weights of the PCs as a function of time, we find for the scattering length $a_\mathrm{s} = 96.65\, a_0$, where we expect to find the isolated Higgs mode, that both the weights of Higgs and breathing patterns (Fig.~\ref{fig:realTimeIsolatedHiggs}(c,$\,$e)) are oscillating at a single frequency of $17.09\pm 0.01\, \mathrm{Hz}$ and $40.46 \pm 0.01\, \mathrm{Hz}$, respectively. The frequencies of the PCs match the BdG results of $21.82\, \mathrm{Hz}$ and $40.50\, \mathrm{Hz}$ at this scattering length quite well, which, in addition to the very similar shape of the modes density variation distributions, underlines that the modes have been correctly identified and that they are excited while being decoupled from each other. The slight difference of the frequencies for the Higgs mode might be due to different numerical grids that were used to carry out the BdG calculation and the full real time evolution. Additionally, we observe a clear envelope of the breathing mode's weight in Fig.~\ref{fig:realTimeIsolatedHiggs}(e), which may indicate that with the ramping procedure, even higher lying modes have been excited. Over the timespan of $300\, \mathrm{ms}$, these higher lying modes decay into the lower lying breathing mode, thus increasing its weight in the timeseries. In contrast, the amplitude of the Higgs weight in Fig.~\ref{fig:realTimeIsolatedHiggs}(c) is approximately constant, indicating that neither the population of this mode increases (no quasiparticle decays into the Higgs mode) nor that it decreases (no quasiparticle is excited from the Higgs to higher lying modes). This can be understood from Fig.~\ref{fig:spectrum}(a), since at $a_\mathrm{s} = 96.65\, a_0$, the Higgs branch is energetically well separated from the more densely distributed, higher lying symmetric modes. Thus, it is less likely to exchange the quasiparticle occupation with any of the higher symmetric modes.

In contrast, for $a_\mathrm{s} = 96.45\, a_0$, where we expect to find the coupled Higgs mode, there are multiple frequency components in the signal, which we can see again from the weights (Fig.~\ref{fig:realTimeCoupledHiggs}(c,$\,$e)). This means that, though we again have clearly excited both the Higgs and breathing mode, they are now coupled to other modes. As a result, the identification of the Higgs mode in this regime is complicated by the fact that despite the main density variation in a timeseries such as Fig.~\ref{fig:realTimeCoupledHiggs}(a) being described by the two PCs in Fig.~\ref{fig:realTimeCoupledHiggs}(b,$\,$d), their weight's frequencies can not be assigned to a single elementary exciation anymore.

Another way to identify the excitation of a specific mode in the experimental protocol, is to reconstruct the signal only using  specific subsets of PCs, which must not necessarily be the ones with highest variance, and compare it to the time evolution one obtains from the BdG equations, as in Fig.~\ref{fig:followinghiggs}(b-g). E.g. if one is interested in whether 2 arbitrary PCs $m_1,\,m_2$ correspond to a specific (or a superposition of) BdG mode(s), one may study the expression $\boldsymbol{M} +  Y_{{m_1}j} \boldsymbol{A}_{m_1} + Y_{{m_2}j} \boldsymbol{A}_{m_2}$ for $j=1,\ ...,\ N_t$ instead of Eq.~\ref{eq:PCAApproxSum}, where $m_1,\,m_2 \neq 1,\,2$. Calculating a measure on how close to the BdG time evolution (Eq.~\ref{eq:timeDensitySlice} in the main text) is to the reconstructed signal using the $m_1$ and $m_2$ PCs can yield better correspondence between BdG and PCA for more complex PCs.

As a final remark, ramping the scattering length further into the isolated droplet regime, $a_\mathrm{s} \lesssim 96\, a_0$, the first PCs do not indicate the spatial shape of the Higgs or breathing pattern as shown in Fig.~\ref{fig:realTimeIsolatedHiggs},\,\ref{fig:realTimeCoupledHiggs}(b,d) anymore, but rather a collection of other highly complex breathing modes. This means that, as expected from the discussion of the decaying Higgs character towards the isolated droplets in the main text, the Higgs mode is not excited by this protocol, or decays into other symmetric modes on timescales that are not accessible in our simulation and in current experimental setups.

\end{document}